\documentclass[aps,prb,twocolumn,superscriptaddress,groupedaddress,amsmath,amssymb]{revtex4-2}
\usepackage{graphicx}
\usepackage[dvipdfmx]{color}
\usepackage{dcolumn}
\usepackage{bm}
\usepackage{color}

\begin{document}

\title{Possible high thermoelectric power factor in alkali-metal-intercalated BC$_3$: anisotropic multiple valleys originating from the van Hove singularity of graphene}
\author{
Ryutaro Enami$^1$,
Kazuhiko Kuroki$^1$, 
and Masayuki Ochi$^{1,2}$ 
}
\affiliation{
$^1$Department of Physics, University of Osaka, Toyonaka, Osaka 560-0043, Japan\\
$^2$Forefront Research Center, University of Osaka, Toyonaka, Osaka 560-0043, Japan
}
\date{\today}
\begin{abstract}
We theoretically investigate the electronic structure of monolayer BC$_3$ and find that it hosts anisotropic multiple valleys originating from the splitting of the van Hove singularity in graphene.
To make use of its favorable electronic structure, we investigate the electronic structure of alkali-metal-intercalated BC$_3$, where intercalated atoms not only introduce electron carriers but also suppress interlayer coupling.
We find that the interlayer transfer is effectively suppressed by potassium intercalation, by which the favorable electronic structure of monolayer BC$_3$ is preserved. 
Finally, we perform model calculation with the onsite-energy offset, and we verify that the strategy of introducing the splitting to the van Hove singularity works well.
\end{abstract}

\maketitle

\section{Introduction}
Thermoelectric energy conversion is an important eco-friendly technology for energy harvesting. Exploring high-performance thermoelectric materials has been an urgent and central task in the study of thermoelectric materials.
For this purpose, researchers have successfully found important concepts for material design, such as low-dimensionality (or anisotropy of the effective mass)~\cite{hicks1993effect, hicks1993thermoelectric, usui2017enhanced}, band convergence (multi-valley band dispersion)~\cite{Pei2011convergence,Lee2020band}, and pudding-mold-shaped band dispersion~\cite{kuroki2007pudding}. In these electronic structures, high density of states (DOS) is realized at the band edge, which is beneficial for efficient thermoelectric conversion.
In fact, many high-performance thermoelectric materials exhibit such characteristics in their electronic structure.
On the other hand, many well-known thermoelectric materials, such as Bi$_2$Te$_3$~\cite{mamur2018review} and PbTe~\cite{heremans2008enhancement}, often contain toxic and/or less abundant elements on the Earth, such as Pb and Te, which is undesirable from an applicational viewpoint.
It is of crucial importance to find high-performance thermoelectric materials consisting of non-toxic and Earth-abundant elements.
Many studies have been conducted along these lines, e.g., on silicides~\cite{nozariasbmarz2017thermoelectric}.
Carbon materials such as (bilayer) graphene and carbon nanotube have also been actively investigated~\cite{mccann2006asymmetry,ohta2006controlling,hao2010thermopower,wang2011enhanced,blackburn2018carbon,horii2021optimization}.

\begin{figure}
    \includegraphics[width=6.5cm]{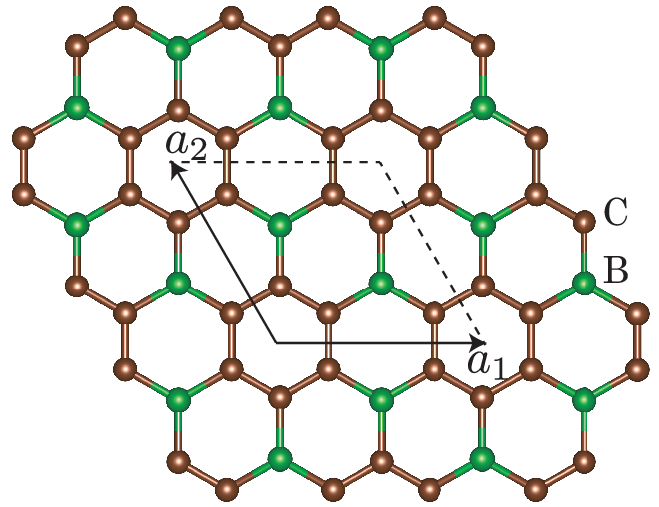}
    \caption{Crystal structure of monolayer BC${}_3$ depicted using the VESTA software~\cite{vesta}. Green and brown spheres represent boron and carbon atoms, respectively. A unit cell is shown with dashed lines.}
    \label{fig:monolayer_struct}
\end{figure}

BC$_3$ is one of the variants of graphene, where carbon atoms in graphene are partially replaced with borons [see, Fig.~\ref{fig:monolayer_struct}]~\cite{kouvetakis1986novel,lee1988electronic,tomanek1988calculation,wentzcovitch1988sigma,kouvetakis1989novel,krishnan1991structure,fecko1993formation,magri1994ordering,miyamoto1994electronic,wang1996stability,wang1997abinitio,jishi2003possibility,sun2004abinitio,tanaka2005novel}.
BC$_3$ has been experimentally synthesized in various forms: bulk~\cite{kouvetakis1986novel,kouvetakis1989novel,krishnan1991structure,fecko1993formation}, monolayer~\cite{tanaka2005novel}, intercalation compounds~\cite{kouvetakis1986novel,kouvetakis1989novel}, and nanotube~\cite{wengsieh1995synthesis,carroll1998effects,fuentes2004formation}.
Due to such diversity in its form, theoretical investigation of BC$_3$ has also been actively conducted not only for bulk and monolayer~\cite{lee1988electronic,tomanek1988calculation,wentzcovitch1988sigma,magri1994ordering,miyamoto1994electronic,wang1996stability,wang1997abinitio,jishi2003possibility,sun2004abinitio,kharabadze2023thermodynamic} but for nanotube~\cite{hernandez1998elastic,jishi1998first,kim2001electronic,kim2001electronic2,fuentes2004formation} and for an alkali-metal-intercalated form as a possible candidate for battery~\cite{ kuzubov2012high,liu2013feasibility,joshi2015hexagonal,kharabadze2023thermodynamic}.
Theoretical investigation on twisted bilayer BC$_3$ is also interesting~\cite{kariyado2023twisted}.
It is noteworthy that the electronic band dispersion is gapped for monolayer BC$_3$~\cite{lee1988electronic,tomanek1988calculation,wentzcovitch1988sigma,miyamoto1994electronic} in contrast to graphene, which makes BC$_3$ a promising candidate for thermoelectric materials.
While the electronic band dispersion and the thermal conductivity~\cite{mortazavi2019outstanding,song2019thermal} have been theoretically investigated, the Seebeck coefficient and the thermoelectric power factor of monolayer BC$_3$ have not yet been explored.
On the other hand, a possible high thermoelectric performance in monolayer C$_3$N was theoretically pointed out in Ref.~\cite{Jiao2022surprisingly}, thus we can expect a high thermoelectric performance also in BC$_3$ because BC$_3$ is a counterpart of C$_3$N in the sense that hole and electron carriers are doped into graphene by boron and nitrogen substitution for carbon, respectively.
However, it is a non-trivial question whether BC$_3$ exhibits high thermoelectric performance.

In this study, we first theoretically investigate the electronic structure of monolayer BC$_3$. We find that monolayer BC$_3$ hosts anisotropic multiple valleys originating from the splitting of the van Hove singularity in graphene, which is caused by the inequivalency between boron and carbon atoms~\cite{note_ineq}.
While this feature of the band dispersion is favorable for thermoelectric materials, it is known that bulk BC$_3$ is metallic~\cite{kouvetakis1986novel} due to the interlayer transfer~\cite{tomanek1988calculation}.
In addition, electron-carrier doping is necessary to employ that band edge. To resolve these problems, we investigate the electronic structure of alkali-metal-intercalated BC$_3$~\cite{kouvetakis1989novel}.
We find that the interlayer transfer is suppressed by potassium intercalation, by which the favorable electronic structure of monolayer BC$_3$ is preserved.
On the other hand, lithium- and sodium-intercalation does not effectively suppress the interlayer transfer.
In addition, electron carriers are doped into the system by alkali-metal intercalation. Considering the suppressed interlayer transfer and carrier concentration tunability, potassium-intercalation is found to be useful to enhance the thermoelectric power factor of BC$_3$.
Finally, we pursue the strategy found here, making use of the anisotropic band edge caused by the split van Hove singularity, in model calculation.
By simple model calculation for the square lattice with the onsite-energy offset, we verify that this strategy enhances the thermoelectric power factor and can expand the possibility of thermoelectric material design.

This paper is organized as follows.
Computational methods used in this study are described in Sec.~\ref{sec:methods}.
First, we discuss the electronic structure of monolayer BC$_3$ and its relation to that of graphene in Sec.~\ref{sec:monolayer}.
Next, we investigate the stable crystal structure, the electronic band dispersion, and the thermoelectric performance of alkali-metal-intercalated BC$_3$ in Sec.~\ref{sec:inter}.
Finally, square-lattice model calculation with the onsite-energy offset is shown in Sec.~\ref{sec:model} to verify our idea that the split van Hove singularity can enhance the thermoelectric performance. Sec.~\ref{sec:summary} summarizes this study.

\section{Methods\label{sec:methods}}
For first-principles calculations, we used the projector augmented wave (PAW) method~\cite{blochl1994projector} and the Perdew-Burke-Ernzerhof parametrization of the generalized gradient approximation (PBE-GGA)~\cite{perdew1996generalized} with the D3 (BJ damping) dispersion correction~\cite{grimme2010consistent,grimme2011effect} as implemented in Vienna {\it ab initio} simulation package~\cite{kresse1996efficient,kresse1993ab,kresse1994ab,kresse1996efficiency}.
A periodic boundary condition was imposed for all calculations shown in this paper.
For B, C, and Na atoms, [He]-core PAW potentials were used.
[Ne]-core PAW potential was used for K, and all electrons were treated as valence for Li.
Both the lattice constants (except the $c$-axis length for graphene and monolayer BC$_3$) and atomic coordinates were optimized until the Hellmann--Feynman force became less than 0.01 eV/\AA\ for each atom. We took $c=15$ \AA\ to eliminate interlayer interaction for graphene and monolayer BC$_3$.
For structural optimization and self-consistent-field calculation, we took $12\times 12\times 1$, $9\times 9\times 1$, $6\times 6\times 6$ ${\bm k}$-meshes for graphene, monolayer BC$_3$, and alkali-metal-intercalated BC$_3$, respectively. DOS for monolayer BC$_3$ was calculated using a $96\times 96 \times 1$ ${\bm k}$-mesh. 
A plane-wave energy cutoff of 500, 500, 700, 850, and 700 eV was taken for graphene, monolayer BC$_3$, Li-intercalated BC$_3$, Na-intercalated BC$_3$, and K-intercalated BC$_3$, respectively.
The optimized crystal structure of monolayer BC$_3$ is shown in Fig.~\ref{fig:monolayer_struct}.
The unit cell of monolayer BC$_3$ contains six C atoms and two B atoms. This unit cell is $2\times 2$ times as large as that for graphene due to the presence of B atoms. 
The C-C and B-C bond lengths for the optimized structure are 1.42 and 1.56 \AA, respectively.

After band-structure calculation, we extracted the Wannier orbitals using Wannier90 software~\cite{pizzi2020wannier90}.
Wannier orbitals were used to obtain the energy plot on a fine ${\bm k}$-mesh and to calculate the transport properties using the tight-binding model.
Here, the tight-binding Hamiltonian is given as
\begin{equation}
\hat{\mathcal{H}} = \sum_{i,j} t_{i,j} {\hat c}^{\dag}_i {\hat c}_j,
\end{equation}
where $t_{i,j}$ and ${\hat c}^{\dag}_i$ (${\hat c}_j$) are the transfer integral between states $i$ and $j$ and the creation (annihilation) operator for an electron of state $i$ ($j$), respectively. The index of the state, $i$ or $j$, specifies the Wannier orbital at each site in the crystal.
Note that the tight-binding model was just used as an interpolation method to obtain the electronic band dispersion on a fine ${\bm k}$-grid. This treatment is valid since only the electronic band dispersion and its derivative are required as system-dependent quantities for calculating the transport coefficient tensors as shown later.
Thus, electrical transport coefficients evaluated with the first-principles band structure and that evaluated with the tight-binding model should coincide if the tight-binding band dispersion well reproduces the first-principles one. We shall see that the agreement is very good in the present system.
For Wannierization, we used $12\times 12\times 1$, $12\times 12\times 1$, and $6\times 6\times 6$ ${\bm k}$-meshes for graphene, monolayer BC$_3$, and alkali-metal-intercalated systems, respectively.
We extracted B-$p_z$ and C-$p_z$ orbitals as Wannier functions for monolayer and alkali-metal-intercalated BC$_3$. C-$p_z$ orbitals were extracted for graphene.

Using the tight-binding model of alkali-metal-intercalated BC$_3$, we evaluated transport coefficients of bulk based on the Boltzmann transport theory. Here, we briefly review the Boltzmann transport theory~\cite{ziman}.
Under the electric field $\mathcal{\bm E}$ and the gradient of the absolute temperature ${\bm \nabla} T$, the Fermi-Dirac distribution function at the local equilibrium $f_{n,{\bm k}}({\bm r}, t)$ for the Kohn-Sham eigenstate for the k-point ${\bm k}$ and the $n$-th band at the position ${\bm r}$ and the time $t$ satisfies
\begin{equation}
\frac{\mathrm{d}f_{n,{\bm k}}}{\mathrm{d}t} = -\frac{f_{n,{\bm k}}({\bm r}, t) - f_{0;n,{\bm k}}}{\tau_{n,{\bm k}}},\label{eq:boltz1}
\end{equation}
where $f_{0;n,{\bm k}}$ is the distribution function at the equilibrium without the external field. Here, we applied the relaxation-time approximation with relaxation time $\tau_{n,{\bm k}}$. By assuming that the system is at the steady state, $\partial f/\partial t = 0$, and that the system is close to the equilibrium since the external fields are weak, $f\simeq f_0$, Eq.~(\ref{eq:boltz1}) can be rewritten as
\begin{equation}
f_{n,{\bm k}}({\bm r}, t) = f_{0;n,{\bm k}} - \tau_{n,{\bm k}} \left( \dot{\bm k}\cdot \frac{\partial f_{0;n,{\bm k}}}{\partial {\bm k}} + \dot{\bm r}\cdot \frac{\partial f_{0;n,{\bm k}}}{\partial {\bm r}} \right). \label{eq:boltz2}
\end{equation}
Semiclassical approximation gives
\begin{equation}
\dot{\bm k} = -\frac{e\mathcal{\bm E}}{\hbar},\ \ 
\dot{\bm r} = {\bm v} {\bm \nabla} T
\end{equation}
with the elemental charge $e$ and the group velocity ${\bm v}$, by which Eq.~(\ref{eq:boltz2}) becomes
\begin{align}
&f_{n,{\bm k}}({\bm r}, t) = f_{0;n,{\bm k}} \notag\\
&+ \tau_{n,{\bm k}} \left( e{\bm v}_{n, {\bm k}} \cdot \mathcal{\bm E} +
\frac{E_{n,{\bm k}}-\mu (T)}{T} {\bm \nabla} {\bm v}_{n, {\bm k}} \cdot T
\right) \frac{\partial f_{0}}{\partial E_{n,{\bm k}}},
\end{align}
where $E_{n,{\bm k}}$ and $\mu (T)$ are the Kohn-Sham eigenenergy for the $n$-th band at the k-point ${\bm k}$ and the chemical potential, respectively.
Then, the current density can be written as
\begin{equation}
{\bm j} = \sum_{n,{\bm k}} (-e {\bm v}_{n, {\bm k}}) f_{n,{\bm k}}({\bm r}, t)
= e^2 K_0 \mathcal{\bm E} + \frac{e}{T}K_1 {\bm \nabla} T,
\end{equation}
with the transport coefficient tensor $K_i$ defined as
\begin{equation}
K_{i} = \sum_{n,{\bm k}} \tau_{n,{\bm k}} {\bm v}_{n,{\bm k}}\otimes{\bm v}_{n,{\bm k}} \bigg[-\frac{\partial f_{0}}{\partial E_{n,{\bm k}}} \bigg] (E_{n,{\bm k}}-\mu(T))^i. \label{eq:transport_coeff}
\end{equation}
Using $K_i$, one can represent the electrical conductivity $\sigma$, the Seebeck coefficient $S$, and the thermoelectric power factor $PF$ as follows:
\begin{align}
\sigma &= e^{2}K_{0} \\
S &= -\frac{K^{-1}_{0}K_{1}}{eT} \\
PF &= \sigma S^{2}.
\end{align}

In our calculation, we first determined the temperature dependence of the chemical potential $\mu(T)$ under the constraint that the number of electrons in the system is preserved. Next, we calculated the transport coefficient tensors, Eq.~(\ref{eq:transport_coeff}), using the eigenstates and eigenvalues of the tight-binding Hamiltonian. Finally, we calculated $\sigma$, $S$, and $PF$ using $K_0$ and $K_1$. All transport calculations were performed under the periodic boundary condition to evaluate bulk transport properties.

In this study, we adopted the constant relaxation-time approximation due to the high computational cost of evaluating the relaxation time in first principle. Instead, we extracted the relaxation time of monolayer C$_3$N from the previous study in a later section just for reference.
Of course, the relaxation time should be different among different materials. In addition, the temperature dependence of $\sigma$ and $PF$ cannot be discussed since the relaxation time, e.g., originating from electron-phonon scattering, strongly depends on the temperature. Thus, first-principles evaluation of $\tau$ is an important future issue.

In this study, we focused on the in-plane transport quantities.
Since the transport properties are isotropic and diagonal (e.g., $\sigma_{xx} = \sigma_{yy}$ and $\sigma_{xy}=\sigma_{yx}=0$) for K-doped BC$_3$ due to $C_{6h}$ symmetry [see Appendix], we shall only show the transport quantities along the $x$ direction ($\sigma_{xx}$ etc.) in the following analysis.
The transport coefficients for the square lattice shown in Sec.~\ref{sec:model} also satisfy, e.g., $\sigma_{xx} = \sigma_{yy}$ and $\sigma_{xy}=\sigma_{yx}=0$, due to the high symmetry of the system.

\section{Results and discussions}

\subsection{Monolayer BC$_3$\label{sec:monolayer}}

\begin{figure}
  \includegraphics[width=8.5cm]{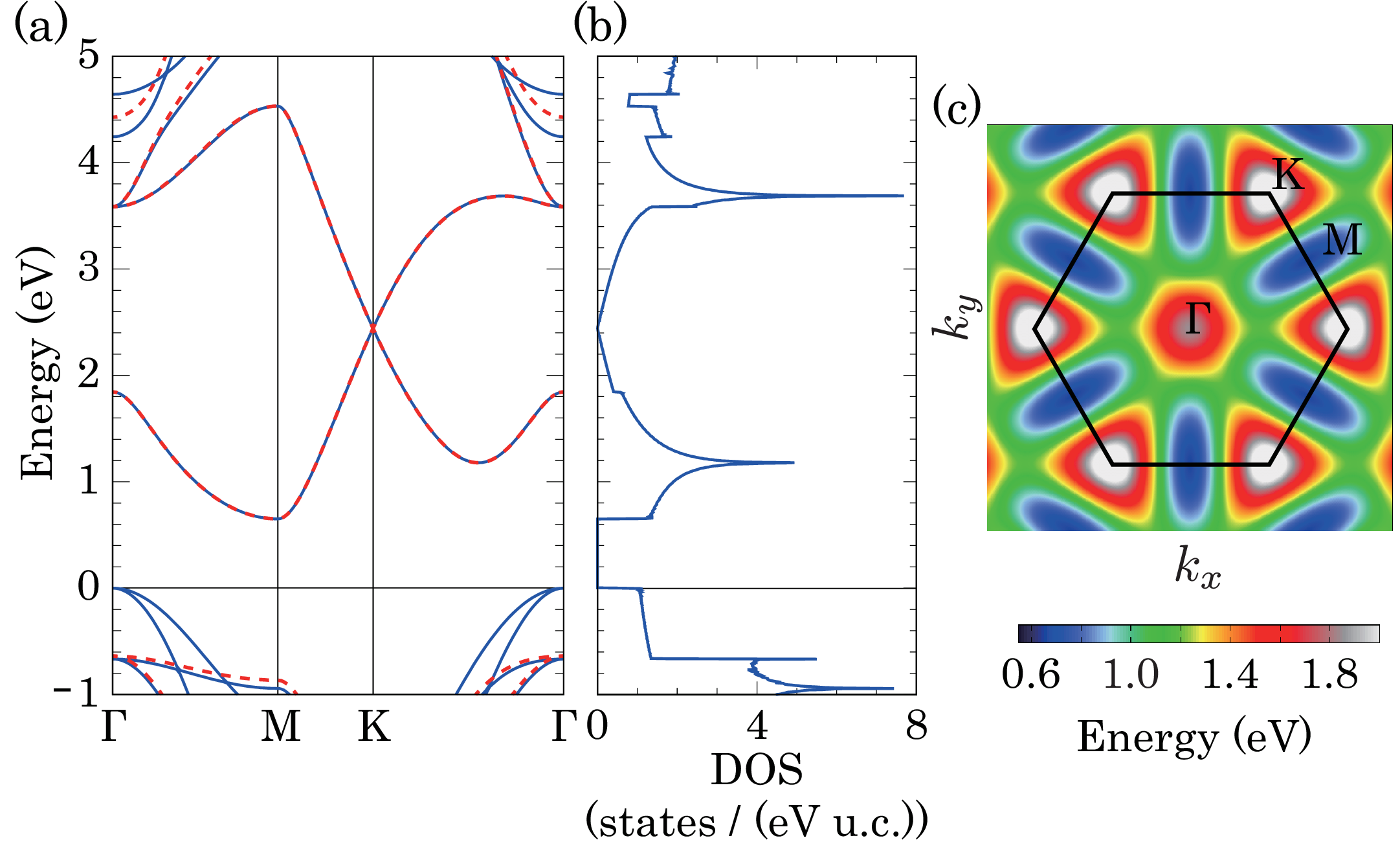}
  \caption{(a) Electronic band structure obtained by first-principles calculation (blue solid lines) and the tight-binding model (red dotted lines), (b) DOS obtained by first-principles calculation, and (c) the lowest conduction band energy on the $(k_x, k_y)$ plane for monolayer BC$_3$ calculated using the tight-binding model. The zero energy is the Fermi energy for all the panels.}
  \label{fig:band_mono}
\end{figure}

By using the optimized crystal structure, we calculated the electronic band structure and DOS of monolayer BC$_3$ as presented in Figs.~\ref{fig:band_mono}(a) and \ref{fig:band_mono}(b), respectively.
As pointed out in previous studies~\cite{lee1988electronic,tomanek1988calculation,wentzcovitch1988sigma}, monolayer BC$_3$ has a band gap in contrast to graphene.
This is an important advantage of BC$_3$ as a thermoelectric material.
In addition, the conduction-band bottom can be regarded as anisotropic multiple valleys, which are also favorable for efficient thermoelectric conversion as described in Introduction.
Here, anisotropy means that the band dispersion around the conduction band bottom at the M point is less dispersive along the $\Gamma$-M line while it is sharp along the M-K line. Note that, while each pocket has an anisotropic shape, $\pm 120^{\circ}$-rotated pockets exist in other M points in the Brillouin zone. These features are clearly illustrated in Fig.~\ref{fig:band_mono}(c), where the lowest conduction band energy is shown on the $(k_x, k_y)$ plane. 
To obtain Fig.~\ref{fig:band_mono}(c) on the fine ${\bm k}$-mesh, we used the tight-binding model extracted from first-principles band structure.
We can clearly see anisotropic electron pockets around three M points, which are colored with blue.
This kind of hidden anisotropy was also found in other materials such as BiS$_2$-based superconductors~\cite{ochi2017prediction}, copper chalcogenides~\cite{ochi2018thermoelectric}, and antiperovskites~\cite{ochi2019comparative}.

\begin{figure}[t]
 \begin{center}
  \includegraphics[width=8.5cm]{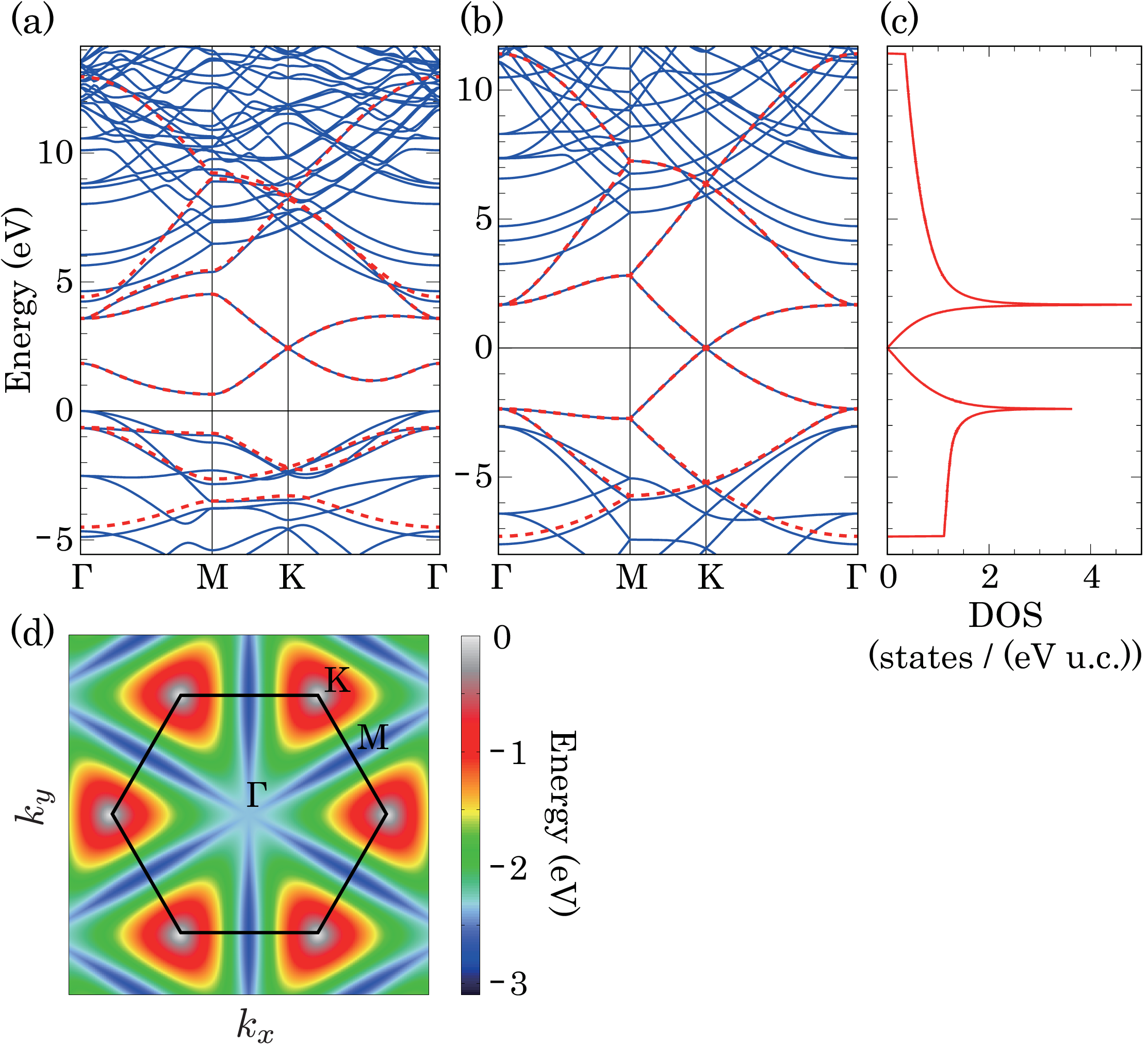}
  \caption{Electronic band structure obtained by first-principles calculation (blue solid lines) and the tight-binding model (red dotted lines) for (a) monolayer BC$_3$ and (b) graphene. The band dispersion of graphene in (b) is $2\times 2$-times folded from the first Brillouin zone of the primitive cell so that the Brillouin zones of monolayer BC$_3$ and graphene are consistent.
  (c) DOS calculated using the C-$p_z$ tight-binding model of graphene. Here, ``per unit cell (u.c.)'' for DOS is defined for the unit cell $2\times 2$-times larger than the primitive one as described above.
  (d) The highest valence band energy in (b) shown on the ($k_x,k_y$) plane calculated using the C-$p_z$ tight-binding model of graphene. The zero energy is the Fermi energy for all the panels.}
  \label{fig:band_graphene}
 \end{center}
\end{figure}

We investigate the origin of the anisotropic multiple valleys in monolayer BC$_3$ by comparing its electronic structure with that for graphene in the following.
Figs.~\ref{fig:band_graphene}(a) and \ref{fig:band_graphene}(b) present the electronic band dispersion of monolayer BC$_3$ and graphene, respectively. The band dispersion of graphene is $2\times 2$-times folded from the first Brillouin zone of the primitive cell so that its Brillouin zone is consistent with that of monolayer BC$_3$.

Although overall band structures are very similar between two compounds, there are two important differences between them.
First, partial substitution of boron for carbon introduces hole carriers into the system, which lowers the Fermi energy shown as zero on the energy axis in Figs.~\ref{fig:band_graphene}(a) and \ref{fig:band_graphene}(b).
Second, the band dispersion is gapped in monolayer BC$_3$ due to the inequivalency between B and C atoms~\cite{wentzcovitch1988sigma}.
We find that the gap opening is caused at the energy of the van Hove singularity of graphene, as verified by DOS of graphene shown in Fig.~\ref{fig:band_graphene}(c).
Here, Fig.~\ref{fig:band_graphene}(c) was calculated using the C-$p_z$ tight-binding model to extract the $p_z$ bands that are mainly relevant to the present discussion.
The isoenergic contour at the van Hove singularity in the valence-band region is shown with blue in Fig.~\ref{fig:band_graphene}(d), where the highest valence band energy is shown on the $(k_x, k_y)$ plane for graphene. An almost flat band dispersion along the $\Gamma$-M line at around $-3$ eV in Fig.~\ref{fig:band_graphene}(b) corresponds to the anisotropic shape of the blue region in Fig.~\ref{fig:band_graphene}(d).
A clear similarity between blue regions in Figs.~\ref{fig:band_mono}(c) and ~\ref{fig:band_graphene}(d) illustrates that the anisotropic multiple valleys of monolayer BC$_3$ originate from the van Hove singularity of graphene.

Another notable issue to be mentioned is that Fig.~\ref{fig:band_graphene}(d) also illustrates that the expected hole carrier concentration for which the Fermi level reaches the van Hove singularity of graphene is consistent with that achieved by C$\to$B substitution in BC$_3$.
In fact, the Fermi pockets around the K points fill up the Brillouin zone at the energy of the van Hove singularity as shown in Fig.~\ref{fig:band_graphene}(d), which means that around two hole carriers per unit cell, including the spin degrees of freedom, are required.
On the other hand, boron substitution introduces two hole carriers in the unit cell of monolayer BC$_3$ since the unit cell includes two boron atoms.
Thus, BC$_3$ is exactly the chemical composition where one can make use of the van Hove singularity of graphene.

\subsection{Alkali-metal-intercalated BC$_3$\label{sec:inter}}

From an applicational viewpoint, it is desirable to find a bulk thermoelectric material rather than a single monolayer. However, it is known that bulk BC$_3$ is metallic due to the band dispersion along the stacking direction~\cite{kouvetakis1986novel, tomanek1988calculation}.
In addition, monolayer BC$_3$ requires carrier doping for use it as a thermoelectric material.

To resolve these problems, we investigate alkali-metal-intercalated BC$_3$.
We expect that interlayer spacing caused by intercalated alkali-metal atoms prevents metallization by reducing the interlayer transfer. In addition, Alkali-metal atoms also introduce electron carriers, by which we can access the favorable electronic structure of the conduction band bottom as discussed in the previous section.

\begin{figure}[t]
 \begin{center}
  \includegraphics[width=8.5cm]{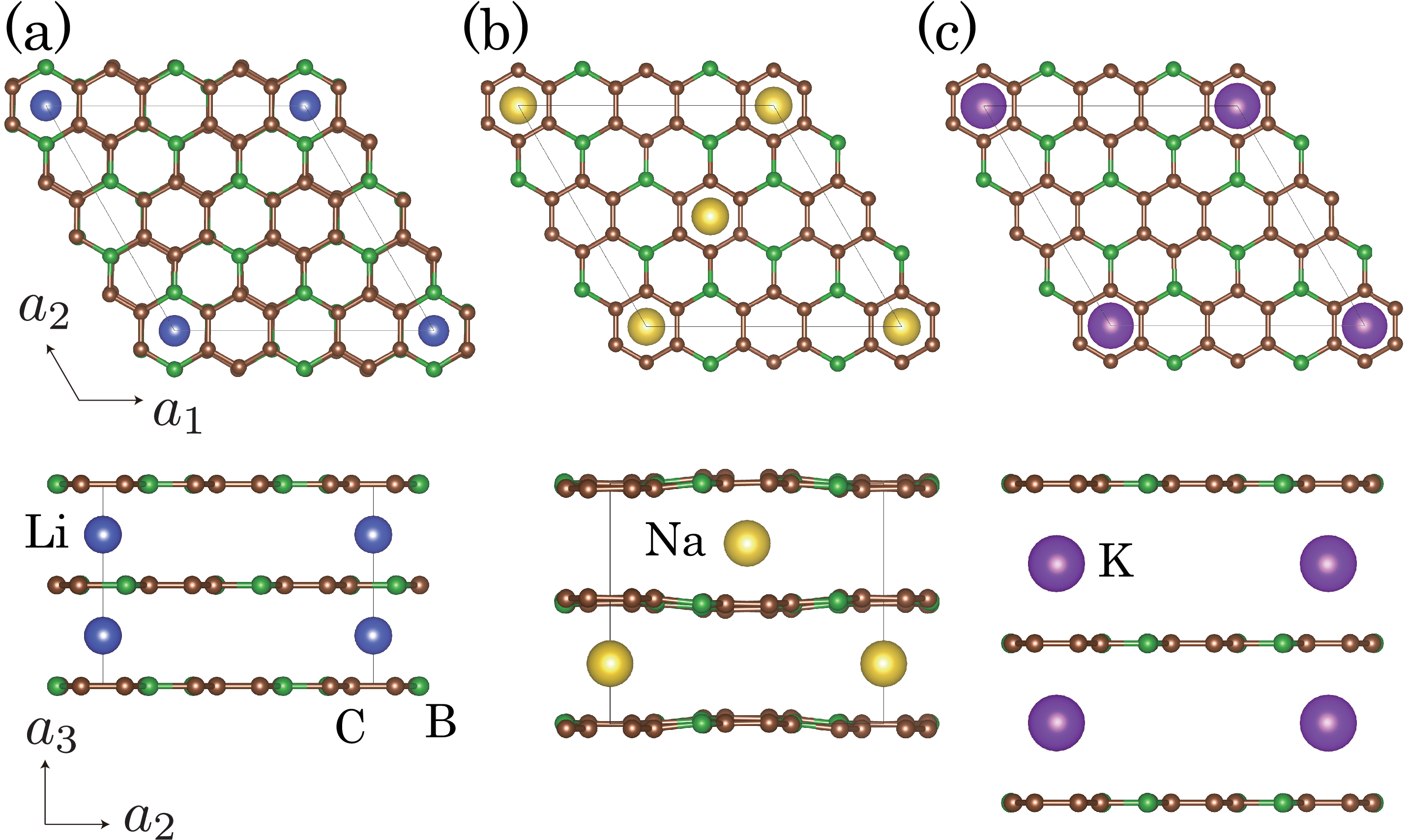}
  \caption{Optimized crystal structures of $X$-intercalated BC$_3$ with $X$ being (a) Li, (b) Na, and (c) K. Upper and lower panels show the top and side views, respectively.
  Green, brown, blue, yellow, and purple spheres represent boron, carbon, lithium, sodium, and potassium atoms, respectively.}
  \label{fig:alkali_struct}
 \end{center}
\end{figure}

In this study, we focused on $X$-intercalated BC$_3$ with $X=$ Li, Na, and K, a synthesis of which was experimentally reported~\cite{kouvetakis1986novel}.
We considered the supercell containing $X_2$B$_{16}$C$_{48}$ where two BC$_3$ layers are included with the AA stacking.
The AA stacking for the bilayer sandwiching the intercalated atoms is a natural assumption often taken for graphite intercalation compounds~\cite{dresselhaus2002intercalation}.
We tried structural optimization of all possible alkali-metal atomic configurations with the restriction that each van der Waals gap has one $X$ atom in the unit cell, and we found the most stable crystal structure for each $X$ as shown in Fig.~\ref{fig:alkali_struct}.
The interlayer distances are 3.34, 3.92, and 5.24 \AA\ for $X=$ Li, Na, and K, respectively.
As naturally expected, a larger atomic radius of the intercalated atom gives a longer interlayer distance.

\begin{figure*}[t]
 \begin{center}
  \includegraphics[width=17.6cm]{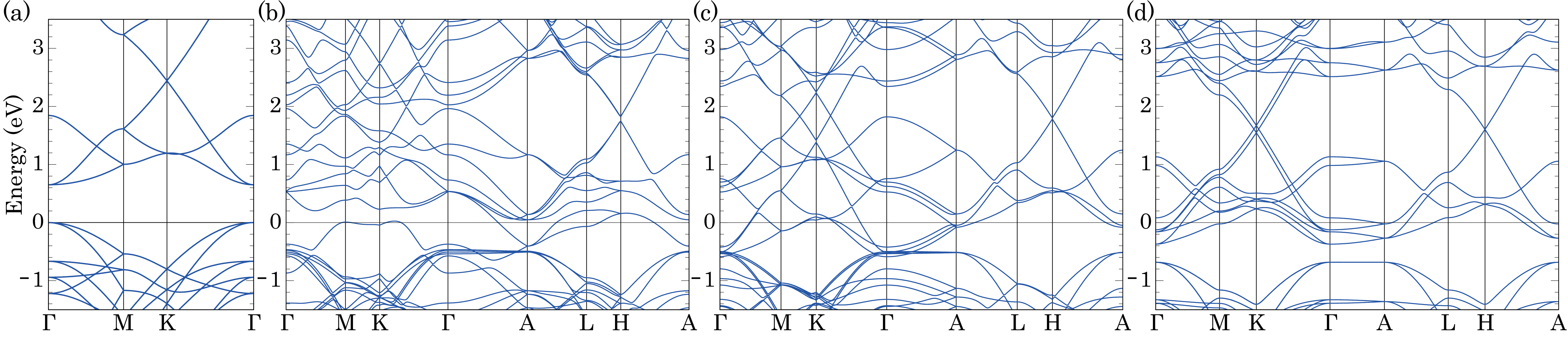}
  \caption{Electronic band dispersion of (a) monolayer, (b) Li-intercalated, (c) Na-intercalated, and (d) K-intercalated BC$_3$. In (a), a folded band structure is shown so that its in-plane periodicity is the same as that for $X$-intercalated BC$_3$. The zero energy is the Fermi energy for all the panels.}
  \label{fig:band_alkali}
 \end{center}
\end{figure*}

The electronic band dispersion of $X$-intercalated BC$_3$ is presented in Fig.~\ref{fig:band_alkali}.
For comparison, a folded monolayer band structure is shown in Fig.~\ref{fig:band_alkali}(a) so that its in-plane periodicity is the same as that for $X$-intercalated BC$_3$.
For $X=$ Li and Na, a relatively large dispersion along the $\Gamma$-A line suggests a large interlayer transfer, by which the system becomes metallic. The original band structure of monolayer BC$_3$ is strongly modified for these cases. On the other hand, K-intercalated BC$_3$ exhibits less-dispersive band structure along the $\Gamma$-A line, and the band structure on the $k_z=0$ plane (i.e., the $\Gamma$-M-K-$\Gamma$ line in the figure) and that for the $k_z=\pi$ plane (i.e., the A-L-H-A line) are similar to that for the monolayer.
As a result, the band dispersion is gapped for K-intercalated BC$_3$.
These results can be naturally understood due to a larger ionic radius of K than Li and Na.

Given the observation made above, we evaluated the thermoelectric power factor for K-intercalated BC$_3$ based on the Boltzmann transport theory with the constant relaxation-time approximation.
For this purpose, we used the tight-binding model consisting of the B-$p_z$ and C-$p_z$ orbitals by extracting Wannier orbitals from the first-principles band structure.
Note that we excluded the valence bands below the band gap for calculating the transport properties.
While the band gap is not very large in Fig.~\ref{fig:band_alkali}(d), a larger band gap is expected in reality since it is well known that PBE-GGA underestimates the band gap. In fact, for monolayer BC$_3$, we verified that the band gap is enlarged by above 1 eV by using the HSE06 functional~\cite{krukau2006influence}.
Thus, we concluded that the bipolar effect by thermally excited hole carriers below the band gap is negligible.

\begin{figure}[t]
 \begin{center}
  \includegraphics[width=7.6cm]{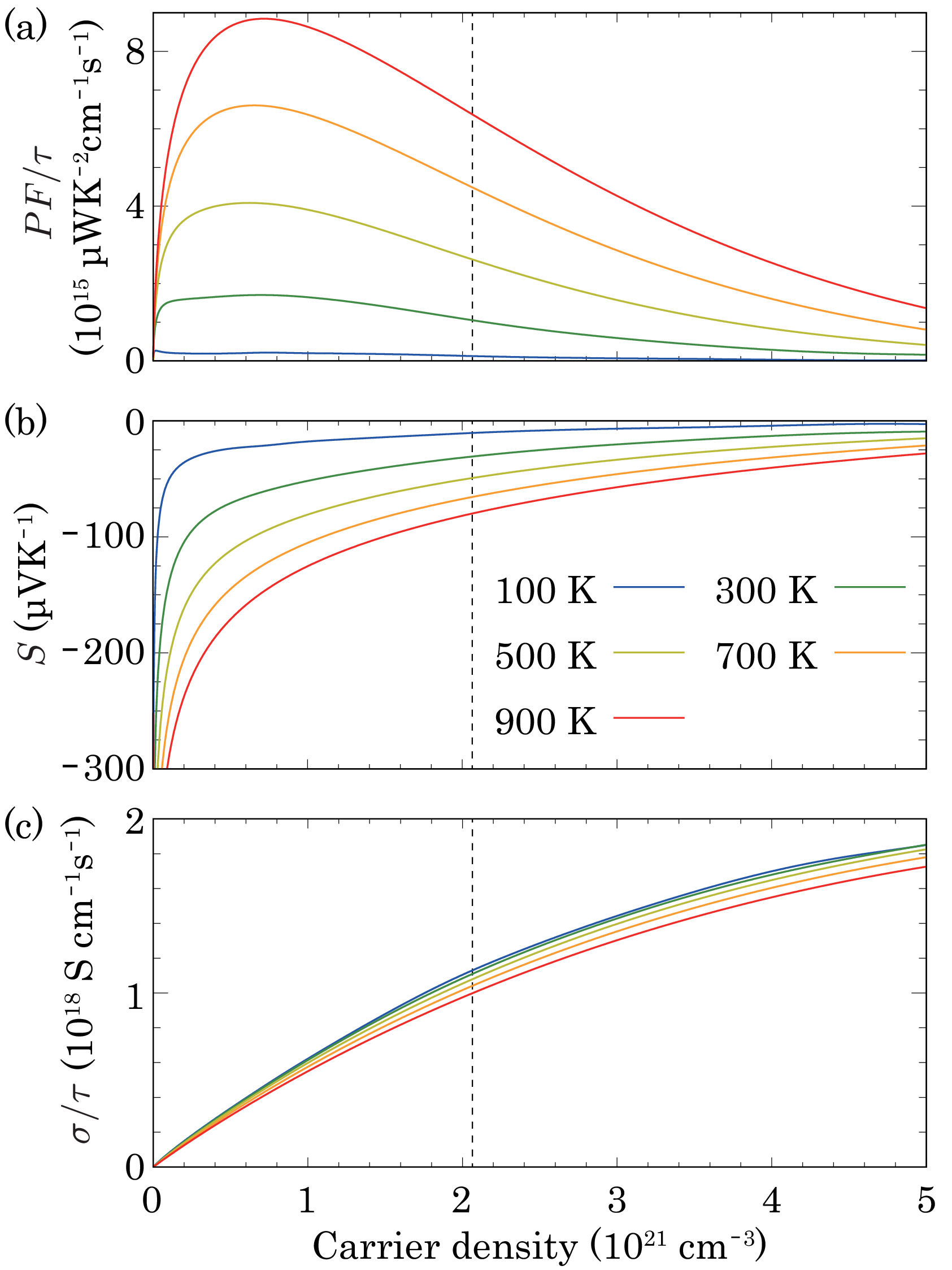}
  \caption{(a) Thermoelectric power factor $PF$, (b) Seebeck coefficient $S$, and (c) electrical conductivity $\sigma$ calculated for K-intercalated BC$_3$ at 100--900 K. For $PF$ and $\sigma$, we show the values divided with the relaxation time $\tau$. Vertical black dotted lines represent the electron carrier density for K$_2$B$_{16}$C$_{48}$ ($=$K$_{0.125}$BC$_3$).}
  \label{fig:thermo_alkali}
 \end{center}
\end{figure}

Fig.~\ref{fig:thermo_alkali} presents transport quantities calculated for K-intercalated BC$_3$ at 100--900 K using a $200\times 200 \times 200$ to $400\times 400 \times 400$ ${\bm k}$-mesh.
Vertical black dotted lines represent the electron carrier density for K$_2$B$_{16}$C$_{48}$ ($=$ K$_{0.125}$BC$_3$).
For more dilute carrier density, $PF/\tau$ reaches a peak of around $2\times 10^{15}\mu$W\ K$^{-2}$cm$^{-1}$s$^{-1}$ at 300 K.
In the previous study~\cite{Jiao2022surprisingly}, the electron relaxation time was evaluated for monolayer C$_3$N, which is a counterpart of BC$_3$ in the sense that hole and electron carriers are doped into graphene by boron and nitrogen substitution for carbon, respectively.
At around $-0.2$ eV from the valence band top in C$_3$N, which corresponds to the $PF$ peak position in K-intercalated BC$_3$, $\tau \sim 1$ -- $5$ fs at 1200 K.
It is well known that the electrical conductivity is inversely proportional to the absolute temperature $T$ at high temperatures. Thus, based on a rough assumption of $\tau \propto T^{-1}$, we can roughly estimate $PF\sim 10$ -- $20\ \mu$W\ K$^{-2}$cm$^{-1}$ for the peak at 300 K. This PF value is comparable to that for high-performance thermoelectric materials~\cite{soleimani2020review}.
First-principles evaluation of $\tau$ for BC$_3$ is computationally expensive but an important future task.
On the other hand, theoretical calculations show that the thermal conductivity of monolayer BC$_3$ amounts to 400--500 W m$^{-1}$K$^{-1}$ at room temperature~\cite{mortazavi2019outstanding,song2019thermal}.
Although this value is an order of magnitude smaller than the thermal conductivity of graphene~\cite{chen2011raman,chen2021thermal,baladin2011thermal}, it is still high for thermoelectric materials. While phonon scattering from intercalated alkali metals might disturb thermal transport to some extent, it is crucial to reduce the thermal conductivity, e.g., through nanostructuring~\cite{hochbaum2008enhanced,Poudel2008high,Boukai2008silicon}, to achieve a high thermoelectric figure of merit.

\subsection{Model calculation for the square lattice\label{sec:model}}

In this paper, we have seen that BC$_3$ can offer a high thermoelectric power factor thanks to its anisotropic multiple valleys originating from the van Hove singularity of graphene.
The van Hove singularity originally existing in graphene is split and then gapped in BC$_3$ due to the inequivalency between boron and carbon atoms.
We can expect that introducing the split of the van Hove singularity can be a good strategy to get promising thermoelectric materials in general situations.
To verify this idea, we performed a simple model calculation for the square lattice.

\begin{figure*}[t]
 \begin{center}
  \includegraphics[width=17.0cm]{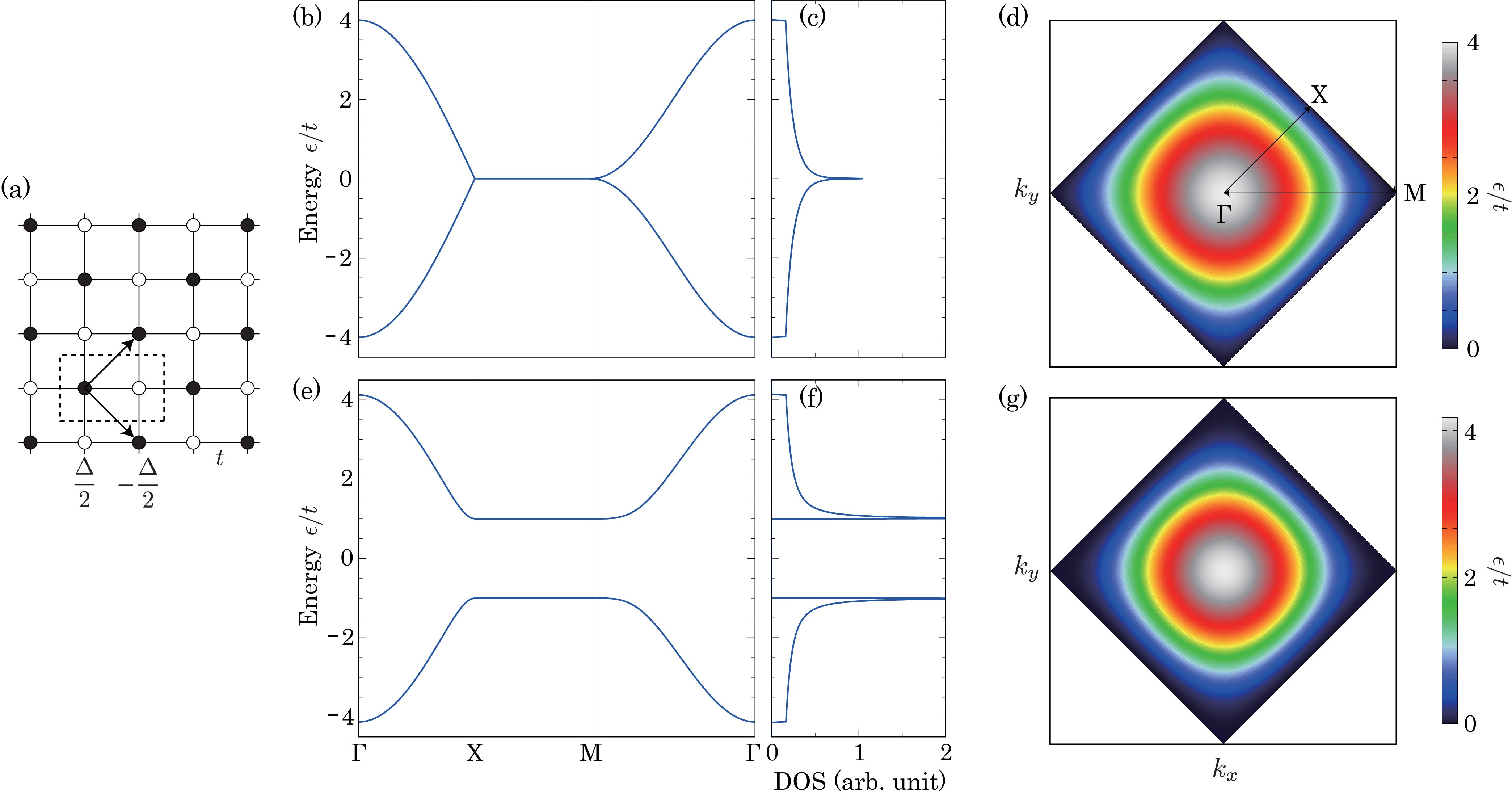}
  \caption{(a) Square lattice model with the onsite energy offset. Dotted lines and arrows present the unit cell and the lattice vectors, respectively. (b) Band dispersion, (c) DOS, and (d) the higher band energy shown on the ($k_x$, $k_y$) plane for model A ($\Delta/t = 0$). (e)--(g) Those for model B ($\Delta/t = 2$).}
  \label{fig:model}
 \end{center}
\end{figure*}

The square-lattice model we investigated here is shown in Fig.~\ref{fig:model}(a).
We considered the nearest-neighbor hopping $t$ and the on-site energy for white and black sites are $\Delta/2$ and $-\Delta/2$, respectively.
We set $\Delta/t=0$ and 2 for models A and B, respectively.
By investigating differences in transport coefficients between two models, we can see the role of band splitting induced by the on-site energy offset $\Delta$. In fact, we will later see that $\Delta$ induces splitting of the van Hove singularity of the square lattice. Thus, the present system is one of the minimal models representing a split van Hove singularity.
The unit cell and the lattice vectors are shown with black dotted lines and black arrows, respectively, in Fig.~\ref{fig:model} (a).

The band structure and DOS for these two models are shown in Figs.~\ref{fig:model} (b)--(g).
While model A has a non-gapped band structure with the van Hove singularity at the center of DOS, the band gap is introduced at the energy of the van Hove singularity in model B due to non-zero $\Delta$.
For model B, the band edges at the energy of $\epsilon/t = \pm 1$ are anisotropic: flat along the X-M line but dispersive along the $\Gamma$-X line. This anisotropic band dispersion can be found at the van Hove singularity, $\epsilon = 0$, for model A.

\begin{figure}[t]
 \begin{center}
  \includegraphics[width=7.0cm]{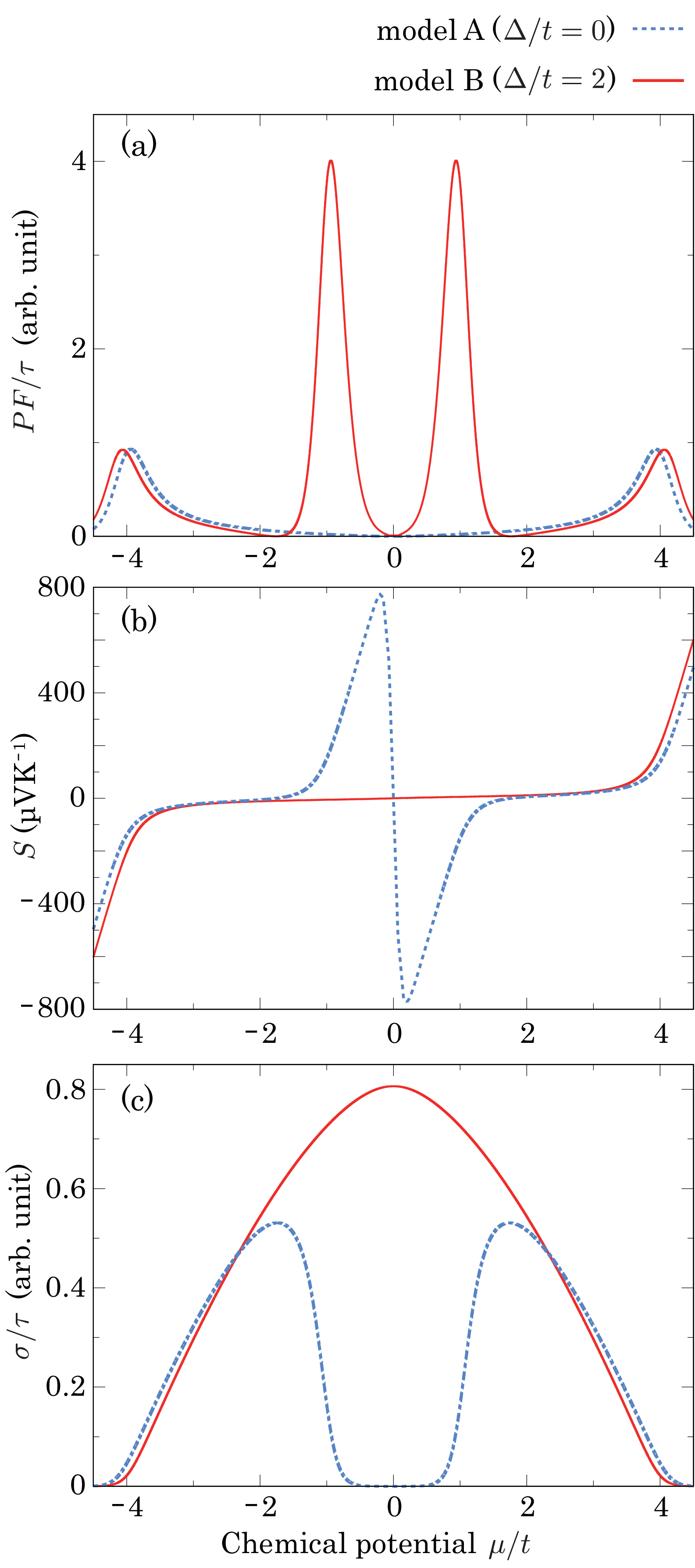}
  \caption{(a) $PF/\tau$, (b) $S$, and (c) $\sigma/\tau$ for models A and B at the temperature $T=0.1 t$. Note that we can set the unit of the Seebeck coefficient even in model calculations since $S=(k_{\mathrm{B}} e^{-1}) (k_{\mathrm{B}}T K_0)^{-1} K_1$ is a product of the constant $k_{\mathrm{B}} e^{-1} = 86\  \mu$VK$^{-1}$ and the dimensionless quantity $(k_{\mathrm{B}}T K_0)^{-1} K_1$, where $k_{\mathrm{B}}$ is the Boltzmann constant.}
  \label{fig:model_thermo}
 \end{center}
\end{figure}

We calculated the thermoelectric power factor $PF$ of these two models as shown in Fig.~\ref{fig:model_thermo}(a).
As a result of the anisotropic band edge originating from the split van Hove singularity, $PF$ has a peak at $\epsilon/t=\pm 1$ for model B.
Those PF peak heights are much higher than those at around $\epsilon/t = \pm 4$ in models A and B.
The calculated Seebeck coefficient $S$ and the electrical conductivity $\sigma$ are also shown in Figs.~\ref{fig:model_thermo}(b)--(c), respectively.
We can see that the enhancement of $PF$ at $\epsilon/t=\pm 1$ in model B is due to a larger $\sigma$ than that around the band edge at $\epsilon/t = \pm 4$, which is expected due to the large DOS at the band gap for model B due to the split van Hove singularity.
Thus, the split of the van Hove singularity introduced by the onsite-energy offset is found to be a good strategy to get high PF.

We expect that the mechanism investigated here can work if one can find the split van Hove singularity since an idea that a high DOS can lead to a high $PF$ has been well established, e.g., in low-dimensional materials and materials having multi-valley band dispersion. While we do not have a specific proposal of real materials other than BC$_3$ at present, one possible way might be starting from the model having a van Hove singularity and then introducing inequivalency among sites (atoms) to open a band gap. A study along these lines is an important future issue.

We also note that electron correlation effects and resulting quantum fluctuations are known to be enhanced by the van Hove singularity in many strongly correlated electron systems. While electron-electron interaction will not play an important role in K-doped BC$_3$, which is a less-localized $p$-orbital system with dilute carrier concentration, the role of electron-electron interaction on systems where $PF$ is enhanced based on our strategy using the van Hove singularity is also an important future issue.

\section{Summary\label{sec:summary}}

We have theoretically investigated the electronic structure of monolayer BC$_3$ and found that monolayer BC$_3$ hosts anisotropic multiple valleys originating from the splitting of the van Hove singularity in graphene.
This splitting is caused by the inequivalency between boron and carbon atoms.
To make use of the favorable electronic structure, we have investigated the electronic structure of alkali-metal-intercalated BC$_3$, where intercalated atoms not only introduce the electron carriers but also suppress the interlayer coupling.
We have found that the interlayer transfer is effectively suppressed by potassium intercalation, by which the favorable electronic structure of monolayer BC$_3$ is preserved. 
In addition, we have performed model calculation with the onsite-energy offset, and verified that the strategy, introducing the splitting to the van Hove singularity, works well.
Our study will expand the possibility of thermoelectric material design.

\section*{Acknowledgment}
This study was supported by JSPS KAKENHI Grants No.~JP22K04908.
The computing resource was supported by the supercomputer system in the Institute for Solid State Physics, the University of Tokyo.

\section*{Appendix: Crystal structures of K-intercalated BC$_3$}

Table~\ref{table:KBC3} presents the optimized crystal structure of K-intercalated BC$_3$. Since the unit cell shown in Fig.~\ref{fig:alkali_struct}(c) can be reduced to a smaller one in terms of the $a_3$-direction, atomic coordinates in a reduced cell of KB$_8$C$_{24}$ are shown in the table.

\begin{table}[t]
\begin{tabular}{ccccc}
\hline
Atom & WP & $x$ & $y$ & $z$ \\
\hline
K & 1b & 0.0 & 0.0 & 1/2 \\
B1 & 2c & 1/3 & 2/3 & 0 \\
B2 & 6l & 0.1674 & 0.3348 & 0.0 \\
C1 & 6l & 0.0801 & 0.1602 & 0.0 \\
C2 & 6l & 0.4205 & 0.8410 & 0.0 \\
C3 & 12p & 0.4207 & 0.3410 & 0.0 \\
\hline
\end{tabular}
\caption{Optimized atomic coordinates of KB$_8$C$_{24}$ with the lattice constants $a=10.33$\ \AA \ and $c=5.24$ \AA .
Wyckoff positions (WP) are shown for the space group $P6/mmm$ (no.~191).
Each atomic coordinate is represented as ${\bm r} = x {\bm a}_1 + y {\bm a}_2 + z {\bm a}_3$ with the lattice vectors ${\bm a}_1 = (a, 0, 0)$, ${\bm a}_2 = (-a/2, \sqrt{3}a/2, 0)$, and ${\bm a}_3 = (0, 0, c)$.}
\label{table:KBC3}
\end{table}

\bibliography{main}

\begin{thebibliography}{68}
\bibitem [{not()}]{note_ineq}%
  \BibitemOpen
  \href@noop {} {}\bibinfo {note} {In our model calculation (not shown in the
  paper), we find that both an onsite energy difference between boron and
  carbon atoms and an inequivalent interatomic transfer energy between B-B and
  B-C bonds are important for causing the split. The former factor, the onsite
  energy difference, comes from the difference in the $2p$ atomic orbital
  energy. The latter factor, the inequivalent interatomic transfer energy,
  correlates with a difference in electronegativity, a size of atomic orbitals,
  and a change in atomic coordinates. In this study, we do not focus on sorting
  out several correlated aspects induced by the inequivalency between boron and
  carbon atoms.}\BibitemShut {Stop}%
\end{thebibliography}
\end{document}